# Shock wave minimized supersonic flight by heating

by David Jonsson, Sweden, david@djk.se,

**Abstract**
Bow shock waves radiate a lot of energy away around supersonic objects in air. To minimize such waves is urgent for fast flying objects. Here is a theoretical description of how shock waves can be prevented by special distribution of air heating.

**Introduction**
For air around a moving air vehicle to change its motion some force has to act on the air. Forces on air are equivalent to pressure gradients $\nabla p = -f$ (1). Pressure changes in air move at the speed of sound. If a vehicle moves faster than the speed of sound the surrounding air can not change its motion as at lower speeds and chock waves form. This shock wave contributes significantly to air drag and is called wave drag. To achieve shock free supersonic flight the forces in the air have to move faster than the speed of sound. Methods to achieve shock free supersonic flight is mentioned by Paul Hill [1] and Semenov et al [2] in various forms. Hill refers to observations of lack of sonic boom of certain vehicles. Below follows an isochoric (constant volume or incompressible) continuous way to explain the lowered wave air drag.

Hill is not specific as to the physical nature of the applied force. He mentions it only as a force field. According to (1) a force field is identical to the gradient of a pressure field. A pressure field in a a gas is also related to temperature. Pressure in air can be ordinary gas kinetic pressure, magnetic pressure or some other form of pressure.

This calculation aims at designing air flow around an object. The flow should be rotational free potential flow. At the air speeds we design for the flow can be considered inviscid (viscosity, thermal conduction and diffusion are ignored) .

To achieve a certain flow forces have to be applied in the fluid. Volumetric forces are equal to pressure gradients (1). Pressure gradients can for example be achieved by heating the air. A simple approximation is to use the ideal gas law pV = nRT or  p = nRT/V=RT/$V_m$ (2) where $V_m$ is the molar volume, which makes in combination with (1) and incompressible flow, since the process is isochoric,

$\frac{R}{V_m}\nabla T = -f$ (3). Forces are applied to change the flow, not to compress or decompress the gas, thus

the assumption of incompressible and isochoric flow holds. The forces around the flying object are conservative. This can in reality be hard to achieve. The second law of thermodynamics says that cooling of the gas can be a problem and loss of energy will occur if the system is designed to have sufficient cooling.

The pressure field from (3) must be used to cause the motion changes in the air. Using m**a** = **F** or per volume ρ**a** = **f** and (1) and (3) gives in the Euler equation

$$\rho a = \rho \frac{D v}{Dt} = \rho\left(\frac{\partial v}{\partial t} + (v \cdot \nabla) v\right) = f = -\nabla p = -\frac{R}{V_m} \nabla T \quad (7)$$

This equation can be solved if we determine the convective derivate D**v**/Dt for the velocity from the desired flow expressed in (6). The time derivative disappears because of steady flow. Together with a vector identity equation 8 is derived.

$$\frac{D v}{Dt} = \frac{\partial v}{\partial t} + (v \cdot \nabla) v = (v \cdot \nabla) v = \nabla v^2/2 - v \times (\nabla \times v) \quad (8)$$

Since we design for potential flow there is no rotation in the fluid $\nabla \times v = 0$ and only the first term of (8) remains which causes (7) to become

$$\rho \nabla v^2/2 = -\frac{R}{V_m} \nabla T \quad (9)$$

or

$$\nabla(\rho v^2/2 + \frac{R}{V_m} T) = 0 \quad (10)$$

or

$$\rho v^2/2 + \frac{R}{V_m} T = constant = C_0 \quad (11)$$

which is the well known Bernoulli's principle with p replaced by (2). $C_0$ can be determined from equation (11) with v = $v_0$ and T = $T_0$ to be $C_0$ = $\rho v_0^2/2 + \frac{R}{V_m} T_0$ , the speed and temperature of the air far from the vehicle. Let us instead use the temperature increase T' = T-$T_0$ to get $C'_0$ = $\rho v_0^2/2$ or

$$\rho(v^2 - v_0^2)/2 + \frac{R}{V_m} T' = 0$$

The desired ideal flow at any speed was determined to be potential flow. It can be calculated for flow around a sphere of radius $r_0$. The flow potential is

$$\Phi = -r_0^3 (v_0 \cdot r)/2r^3 \quad (4)$$

and the flow

$$v_1 = \nabla \Phi = -r_0^3 v_0/2r^3 + 3 r_0^3 (v_0 \cdot r) r/2r^5 \quad (5)$$

or as seen in steady flow around a sphere at rest.

$$v = -v_0 - r_0^3 v_0/2r^3 + 3 r_0^3 (v_0 \cdot r) r/2r^5 \quad (6)$$

The desired temperature distribution is thus determined by using expression (6) for **v**.

$$T' = \frac{\rho V_m (v_0^2 - v^2)}{2R} = \frac{\rho V_m v_0^2}{2R}\left(1 - \sin^2\theta\left(1 + \frac{r_0^3}{2r^3}\right)^2 - \cos^2\theta\left(1 - \frac{r_0^3}{r^3}\right)^2\right) \quad (12)$$

This equation is visualized in figure 1 and 2.

The many observed phenomena of lowered drag by heating can thus be explained. The more the heat distribution of an experiment or observed shock free vehicle resembles equation (12) the less drag it will experience.

**Physical feasibility of air heating**
The maximum heat will be at the end and rear of the sphere. The maximum heat for the speed of sound in dry air at STP.

$$max(T') = \frac{\rho V_m v_0^2}{2R} = \frac{1.29 \cdot 0.0248 \cdot 331^2}{2 \cdot 8.31} K = 210 K \quad (13)$$

This is not an unreasonably high temperature. A simple rule would be T' = 210 $M^2$ where M is the Mach number. The minimum temperature is at the midsection where min(T') = -1.25 max(T') = - 262 K which seems very difficult to achieve. To determine the net total heating is important in order to determine efficiency of the heating process and an eventual power saving of the propulsion of the vehicle. As mentioned previously there would be no net heating and thus no power requirement in an ideal process. This seems to be unfeasible because of convection and problems of cooling. Another simple assumption is to assume that cooling of the gas takes place according to some loss function. Such cooling could be so high that heating requirements become excessive or cooling could be so slow that the cooling along a streamline becomes too weak.

To heat in certain spatial distributions is conditioned by physical factors. Heating can traditionally be done by streaming, conduction and radiation. Conduction is considered too low, and was also ruled out by the assumption of inviscid flow. Streaming definitely takes place and is of a known form, the potential flow. Radiation is another way to heat air. Radiation can be done from point or line sources and their combinations. For our simple spherical case we can use a point source. Point source radiation decrease as $1/r^2$. Air is transparent up to the ultraviolet (UV) region of 186 nm. In the UV region the absorption varies significantly and quickly reaches high values with shorter wavelengths. UV radiation ionizes the air. Ionized air will in turn lower transparency for longer wavelength radiation thus making heating with electromagnetic waves possible even with radiation of longer wavelength than 186 nm.

**Distribution of air heating**
The heat around the sphere is stationary seen from the sphere. This means the total or convective derivative is zero from that frame of reference.

$$\frac{DT}{Dt} = \frac{\partial T}{\partial t} + \mathbf{v} \cdot \nabla T = 0 \quad (13)$$

the heating thus becomes

$$\frac{\partial T}{\partial t} = -\mathbf{v} \cdot \nabla T \quad (14)$$

which in the case for the sphere takes the following expression

$$\frac{\partial T}{\partial t} = -\frac{\rho V_m v_0^3}{2R}\left(\frac{3 r_0^3 \cos\theta}{8 r^{10}}\left(-4 r^6 - 16 r_0^3 r^3 + 11 r_0^6 + \cos(2\theta)\left(20 r^6 - 16 r_0^3 r^3 + 5 r_0^6\right)\right)\right) \quad (15)$$

The distribution of heat in the line forward of the direction of flight is plotted in figure 3. The total distribution in a plane parallel with the flight direction is described in a density plot in figure 4.

If this spatial distribution is physically feasible remains to be determined. The sphere is usually not a preferable shape of flying objects and another way to achieve a suitable heating is to determine the shape based on possible distributions of heating. The heating from radiating sources are usually (if not always) distributed as exponentially decreasing.

**Power requirements for air heating**
Equation (15) determines the heating of the surrounding air. It is symmetric around the axis in direction of flight and antisymmetric on the two halves separated by the midsection plane. It would thus be meaningful to integrate the total heating on each of the halves to get an understanding of the required power to heat and cool the flying thing. This power requirement must be compared with the power losses of the wave drag to see if the shock prevention technique is worth applying.

**References**
1. Paul R. Hill, *Unconventional Flying Objects: a scientific analysis*, 1995, Hampton Roads Publishing Co., ISBN 1-57174-027-9
2. V E Semenov et al *2002 Plasma Phys. Control. Fusion* **44** B293-B305 doi:10.1088/0741-3335/44/12B/321 http://www.iop.org/EJ/abstract/0741-3335/44/12B/321/

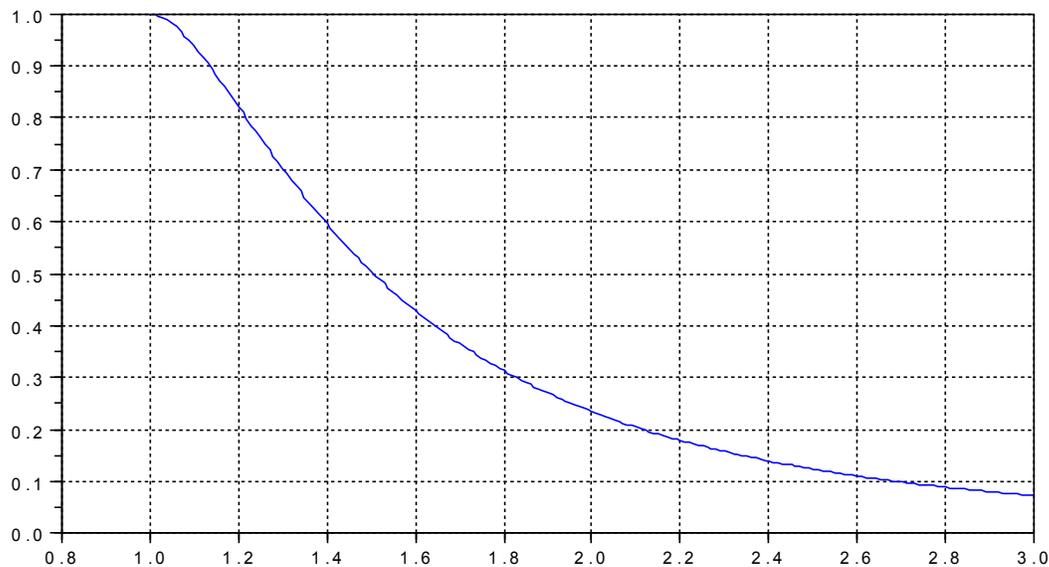

*Figure 1. Relative radial distribution of heat in front and rear of a sphere. Horizontal axis in radii.*

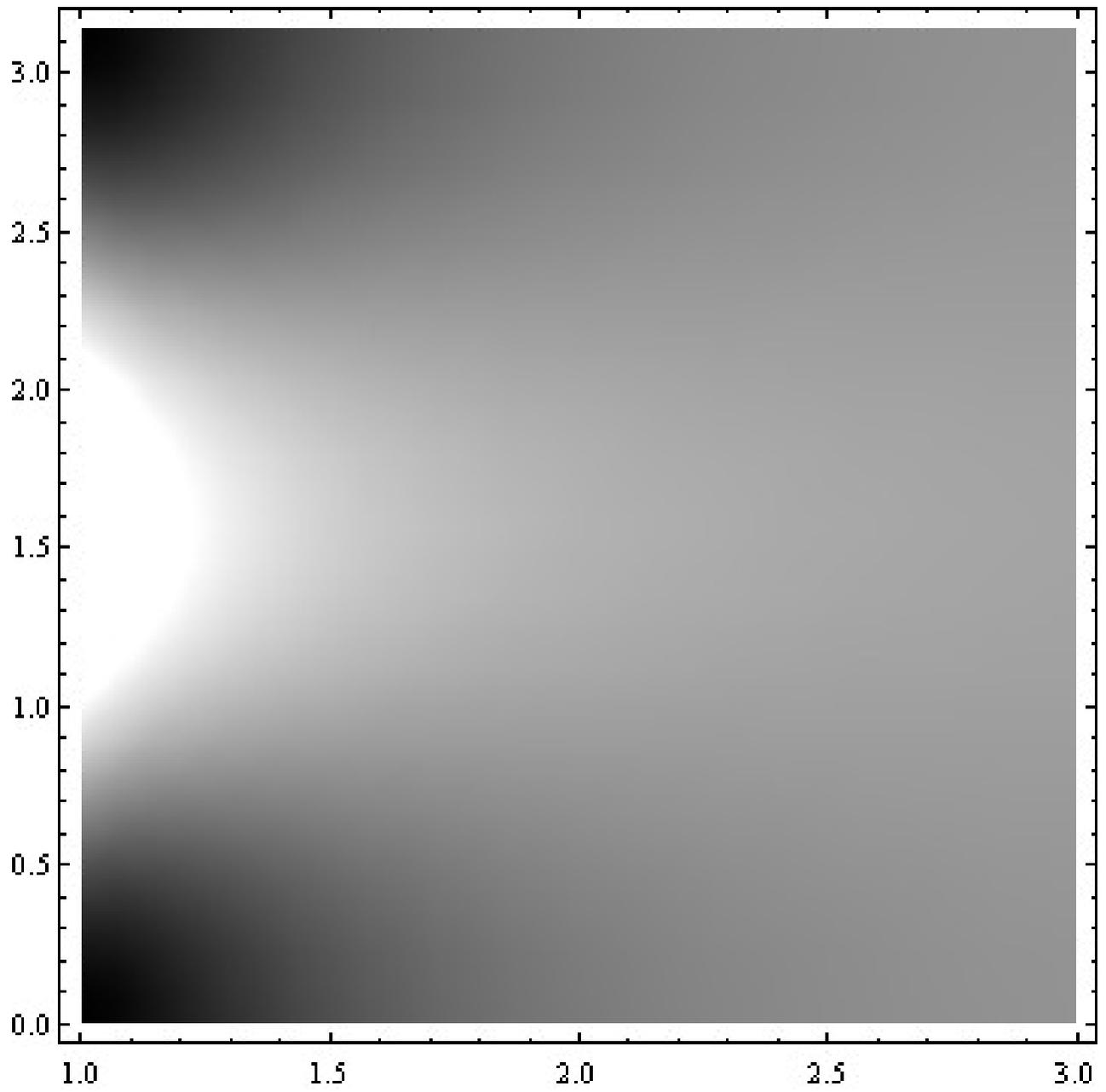

*Figure 2. Temperature around sphere, vertical axis in radians, horizontal axis in radii. Darker is hotter. Motion is downwards.*

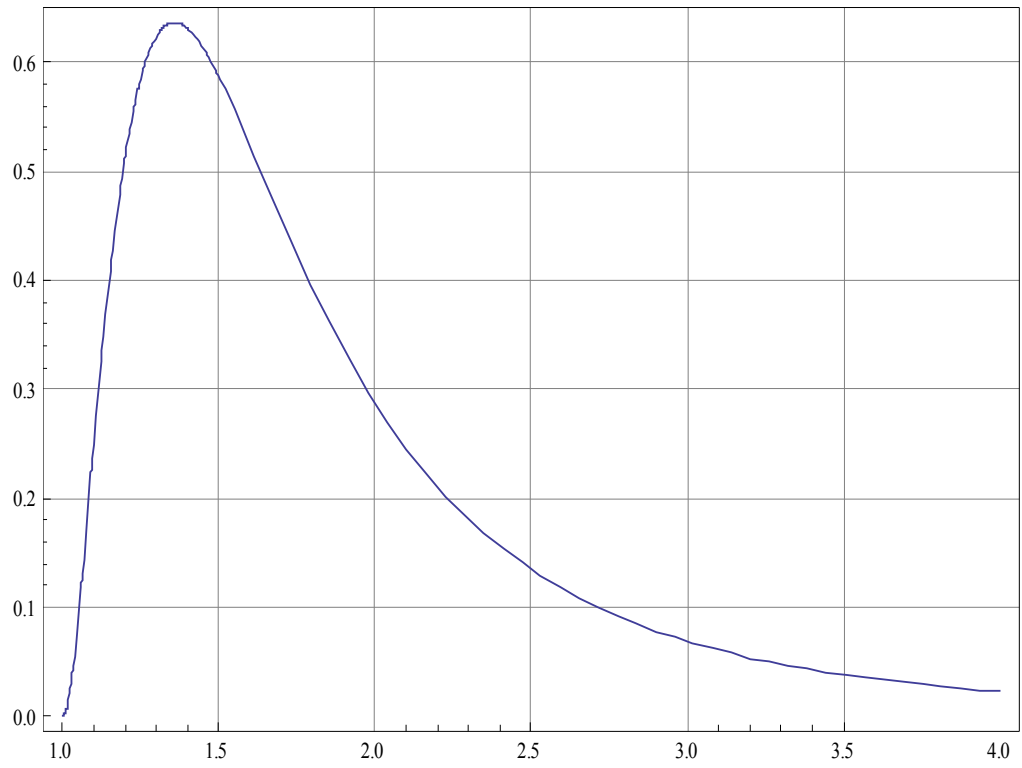

*Figure 3. Relative radial distribution of heating in front of a sphere. Horizontal axis in radii.*

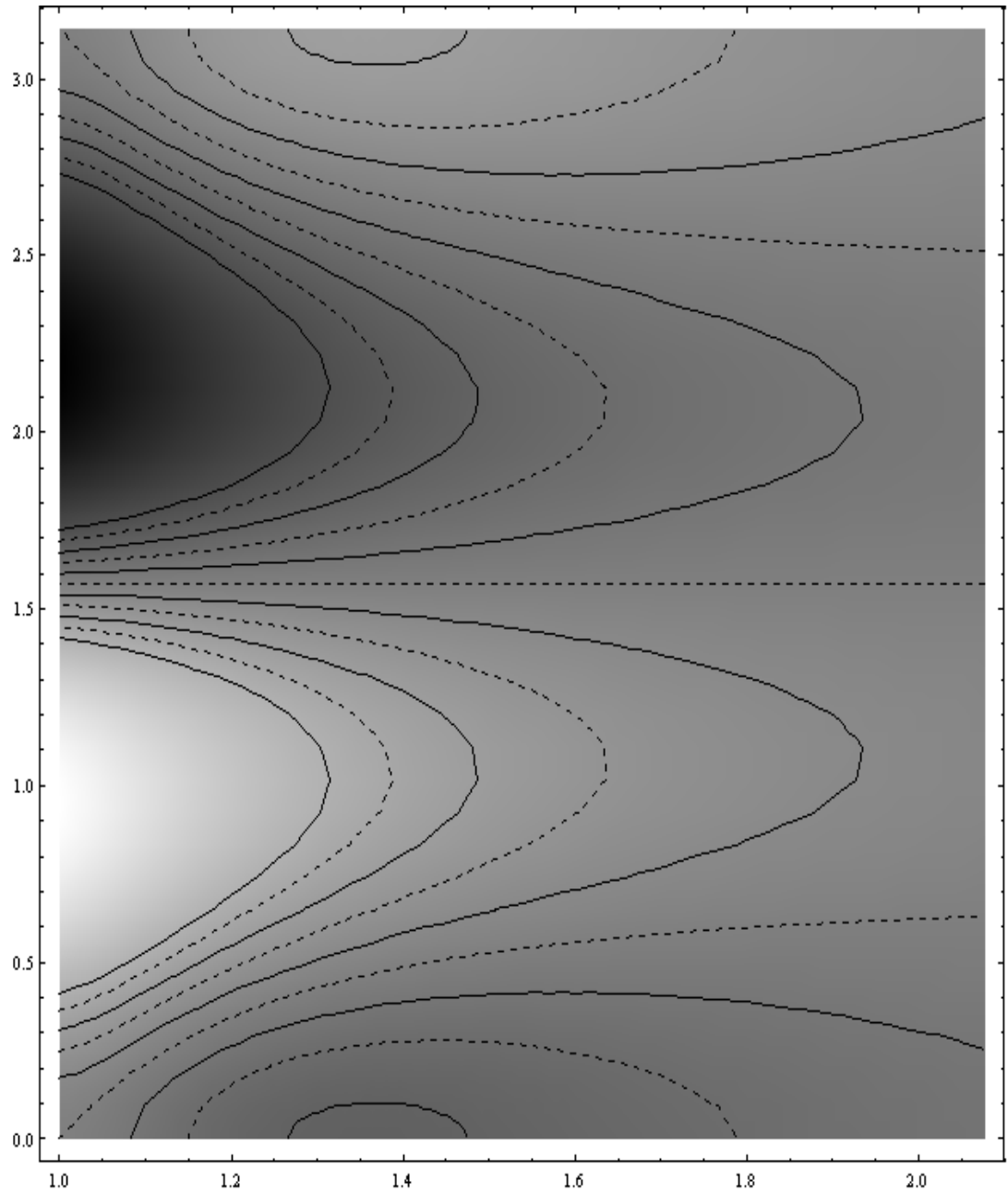

*Figure 4. Heating around sphere, vertical axis in radians, horizontal axis in radii. Whiter is cooled. Darker is heated. Motion is downwards.*